\documentclass[doublecol]{epl} 
\let\oldmarginpar\marginpar
\renewcommand\marginpar[1]{\-\oldmarginpar[\raggedleft\footnotesize #1]%
{\raggedright\footnotesize #1}}

\usepackage[dvipsnames]{xcolor}   
  
\usepackage[normalem]{ulem}		

\usepackage{bm}		
\usepackage{amssymb,amsmath}
\usepackage{graphicx}
\usepackage{comment}

\usepackage{enumitem}   

\usepackage{epstopdf} 
\usepackage{physics}	

\DeclareMathOperator {\e}{e}	
\newcommand{\sbb}{\textrm{sb}}  

\title{Erasing a majority-logic bit}

\author{Karel Proesmans \inst{1,2,3} \and John Bechhoefer \inst{3}}
\institute{\inst{1}Department of Physics, University of Luxembourg, Luxembourg \\\inst{2}Hasselt University, B-3590 Diepenbeek, Belgium\\
\inst{3}Department of Physics, Simon Fraser University, Burnaby, B.C., V5A 1S6, Canada}


\abstract{We study finite-time bit erasure in the context of majority-logic decoding. In particular, we calculate the minimum amount of work needed to erase a majority-logic bit when one has full control over the system dynamics.  Although a single unit bit is easier to erase in the slow-driving limit, the majority-logic bit outperforms the single unit bit in the fast-erasure limit. Our results also suggest optimal design principles for majority-logic bits under limited control.}

\begin{document}

\maketitle

\section{Introduction}
    In his seminal 1961 paper, Landauer showed that there is a fundamental thermodynamic cost associated with the erasure of a bit \cite{landauer1961irreversibility}. In particular, the amount of work needed to erase a bit is always {greater} than $k_\textrm{B}T\ln 2$, where $k_\textrm{B}$ is the Boltzmann constant and $T$ the temperature of the environment. Although modern computers still dissipate several orders of magnitude more than the Landauer limit, Koomey's law predicts that computers working near the Landauer limit will become available over the next couple of decades \cite{koomey2010implications}. This raises  questions about how to optimize the erasure of bits and whether there are any thermodynamic bounds on bit erasure beyond the Landauer limit.

The last two decades have seen a surge in studies on the thermodynamics of bit erasure. On the theoretical side, the framework of stochastic thermodynamics \cite{seifert2012stochastic,van2015ensemble} has enabled the analysis of specific models and led to general predictions concerning the thermodynamics of finite-time bit erasure \cite{dillenschneider2009memory,aurell2012refined,diana2013finite,zulkowski2014optimal,zulkowski2015optimal,deshpande2017designing,melbourne2018realizing,boyd2018shortcuts,riechers2019balancing,proesmans2020finite,proesmans2020optimal}. Meanwhile, Landauer's principle has been verified on a broad class of experimental systems \cite{berut2012experimental,berut2013detailed,jun2014high,gavrilov2016erasure,hong2016experimental,martini2016experimental,gavrilov2017direct,gaudenzi2018quantum,ribezzi2019large,saira2020nonequilibrium,cetiner2020dissipation}.
One general prediction is that the extra work needed to erase a bit over a finite amount of time is inversely proportional to the duration of the protocol, where the proportionality constant depends on the level of control that one has over the system \cite{aurell2011optimal,aurell2012refined,zulkowski2014optimal,zulkowski2015optimal,proesmans2020finite,proesmans2020optimal,schmiedl2008efficiency}.

The focus of these works have generally been on single-unit (SU) bits, i.e., systems where the state of the bit is determined by a single spin or the position of a single colloidal particle. Realistic computers, however, are generally based on majority logic (ML) decoding \cite{richter1999recent}. An ML bit consists of several sub-bits, each of which can be either in state $0$ or $1$. The ML bit is then in state $0$ or $1$ if the majority of sub-bits is in state $0$ or $1$ respectively. 

Inspired by a case study by Sheng et al.~\cite{sheng2019thermodynamics}, we present here a thorough analysis of the thermodynamics of majority-logic decoding. To do so,  we consider an ML bit with a large number of sub-bits under full control. We will show that SU bits are easier to erase in the slow-erasure limit but that ML bits are easier in the fast-erasure limit.  In particular, the work required to erase an ML bit goes as $\tau^{-1/2}$ for fast protocols of duration $\tau$, while an SU bit requires an amount of work proportional to $\tau^{-1}$ under similar conditions.


\section{Stochastic thermodynamics of bit erasure \label{sec:st}}
An SU bit can generally be described by some microscopic variable $x$. One can, for example, think about a superconducting flux, the position of a particle or the magnetization of a spin. We define the bit to be in state $0$ if $x<0$ and in state $1$ if $x>0$. Throughout this paper, we will assume that $x$ is a stochastic quantity that can be described by a probability distribution $p(x,t)$ at time $t$, evolving via a one-dimensional, overdamped Fokker-Planck equation,
\begin{equation}
    \pdv{t} p(x,t)=\frac{D}{k_\textrm{B}T} \pdv{x} \left(p(x,t)\pdv{x}V(x,t)\right) +
        D\frac{\partial^2}{\partial x^2}p(x,t) \,.
\label{eq:FP}
\end{equation}
Here, $V(x,t)$ is the potential energy landscape, $D$ is the diffusion coefficient associated with $x$, $k_\textrm{B}$ is the Boltzmann constant and $T$ is the temperature of the environment. Although we will focus in this paper on one-dimensional overdamped systems, we expect similar results to hold for underdamped and higher-dimensional systems.  Furthermore, we will assume that the potential is initially symmetric, $V(-x,0)=V(x,0)$ and that the system is initially in equilibrium with the potential,
\begin{equation}
    p(x,0) = \pi(x) \equiv \frac{\e^{-\frac{V(x,0)}{k_\textrm{B}T}}}{\int^{\infty}_{-\infty} \dd{y}  \e^{-\frac{V(y,0)}{k_\textrm{B}T}}} \,.
\end{equation}
Thus, at $t=0$, the probability for the bit to be in state $0$, $P_0$, equals the probability for it to be in state $1$, $P_1$. 

Next, we review the design of protocols that erase an SU bit to state $0$ over an amount of time $\tau$ with an erasure error $\epsilon$ \cite{proesmans2020optimal}; i.e., the probability for the bit to be in the wrong state is $\epsilon$. The final distribution $p_\tau(x)$ should then satisfy $P_0 \equiv p(x<0,\tau)=1-\epsilon$ and $P_1 \equiv p(x>0,\tau) = \epsilon$. To find such protocols, we assume we have full control over the potential $V(x,t)$ for $0<t<\tau$. The time-dependent function $V(x,t)$ then defines the protocol.  As a further requirement, the potential-energy landscape should return to its original form at the end of the (cyclic)  protocol, $V(x,\tau)=V(x,0)$.  The work, on average, needed to complete this process has two contributions, 
\begin{align}
    W&=\Delta\mathcal{F}+T\Delta_\textrm{i} S \,,
\label{eq:W}
\end{align}
where  $\Delta \mathcal{F}$ is the non-equilibrium free energy difference between $p_0(x)$ and $p_\tau(x)$, and $\Delta_\textrm{i} S$ is the amount of entropy produced during the protocol.

Using ideas from stochastic thermodynamics and requiring cyclic protocols, one can show that $\Delta \mathcal{F}$ is given by the Kullback–Leibler divergence between the initial and final states \cite{Esposito2011NonequilibriumFreeEnergy},
\begin{equation}
    \Delta \mathcal{F}=k_\textrm{B}T \int^\infty_{-\infty} \dd{x} 
        p(x,\tau) \ln \frac{p(x,\tau)}{p(x,0)} \,.\label{eq:df}
\end{equation}
The entropy production $\Delta_\textrm{i} S$  to transform a state $p(x,0)$ to a state $p(x,\tau)$ in a given time $\tau$ generally depends on the protocol, $V(x,t)$.  Its minimum value, however, can be shown to be given by \cite{aurell2011optimal,aurell2012refined,proesmans2020finite,proesmans2020optimal,zhang2019work,zhang2020optimization}
\begin{equation}
    \Delta_\textrm{i}S_{\textrm{min;SU}} = k_\textrm{B}\frac{\int_0^1 \dd{y} \left(f_0^{-1}(y)-f_{\tau}^{-1}(y)\right)^2 }{D\tau} \,,\label{eq:ds}
\end{equation}
with $y=f_{0/\tau}(x)=\int_{-\infty}^x \dd{x'} p_{0/\tau}(x')$ the cumulative distribution function of the initial/final probability distribution. 

Equation~\ref{eq:W}, together with Eqs.~\eqref{eq:df}--\eqref{eq:ds} gives the least average work to erase a bit given initial and final distributions.  One can further reduce the required work by relaxing the assumption that $p_\tau(x)$ be specified in advance.  It is enough that $p_\tau(x)$ satisfy $f_\tau(0)=1-\epsilon$. Although minimizing $W$ over these allowed final distributions cannot generally  be done analytically, one can show that in the fast-erasure limit, $\tau\rightarrow 0$, the minimum work $W_\textrm{min;SU}$ is \cite{proesmans2020finite}
\begin{equation}
    W_\textrm{min;SU} = \frac{k_\textrm{B}T}{D\tau}\int_0^{f_0^{-1}(1-\epsilon)}\dd{x} 
    p_0(x) x^2 + o\left(\frac{1}{\tau}\right) \,,
\label{eq:fast}
\end{equation}
while in the slow-erasure limit
\begin{equation}
    W_\textrm{min;SU} = k_\textrm{B}T \left[ \epsilon \ln (2\epsilon) + (1-\epsilon) \ln (2(1-\epsilon)) \right]
    +\order{\frac{1}{\tau}} \,.
\label{eq:slowbound}
\end{equation}
Note that Eq.~\eqref{eq:slowbound} reduces to the Landauer bound in the zero-error limit, $\epsilon=0$.

\begin{figure}
    \centering
    \includegraphics[width=3.25in]{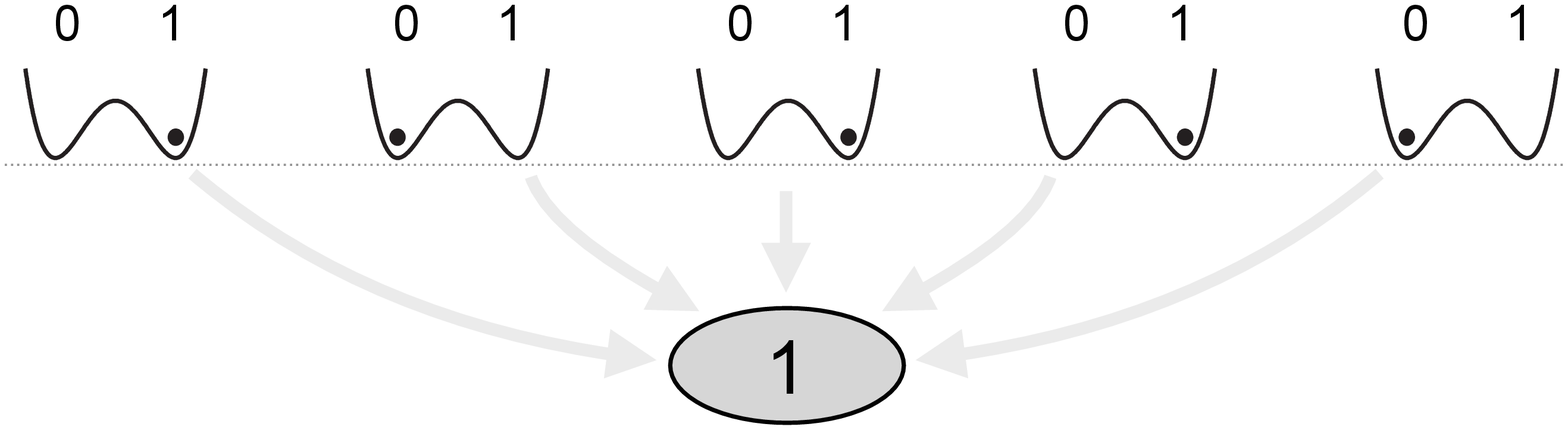}
    \caption{Schematic diagram illustrating how $N=5$ sub-bits combine to form one majority-logic (ML) bit.}
\label{fig:schematic}
\end{figure}

\section{Majority-logic decoding\label{sec:mld}}

Figure~\ref{fig:schematic} shows how the state of an ML bit is determined by the sum of $N$ sub-bits. We will assume that the sub-bits do not interact with each other and that each can be described as an SU bit that obeys Eq.~\eqref{eq:FP}. Since all sub-bits are equivalent, the ML bit is in state $0$ if more than half of the sub-bits are in state $0$ and in state $1$ if more than half are in state $1$. Throughout this paper, we will assume that $N$ is odd, to avoid the possibility of a tie in deciding which state has a majority. If the probability for each sub-bit to be in state $0$ is $1/2+\delta$, one can show that the ML bit is in state $0$ with probability \cite{sheng2019thermodynamics}
\begin{equation}
    P_0 = \frac{1}{2} + \frac{\int_0^{\frac{1}{2}+\delta}\dd{x} (x(1-x))^{\frac{N-1}{2}}}
        {\int_0^{1} \dd{x} (x(1-x))^{\frac{N-1}{2}}} \,.
\label{eq:binomial}
\end{equation}
We show in the appendix \ref{sec:app1} that, for large $N$ and letting $\sqrt{N}\delta \sim \order{1}$, a saddle-point approximation gives
\begin{equation}
    P_0\approx\frac{1}{2} \left( 1+\erf(\sqrt{2N}\delta) \right) \,.
\end{equation}

Thus, to erase an ML bit with an erasure error $\epsilon$, we need to erase all sub-bits by an amount $\delta$, given by
\begin{equation}
    \delta \approx
        \frac{\erf^{-1}(1-2\epsilon)} {\sqrt{2N}} \,.
\end{equation}
Note that $\delta=0$ corresponds to no erasure, as the probabilities to be in state $0$ and $1$ remain equal. 

The total work associated with the erasure of an ML bit with a large number of sub-bits now equals $N$ times the thermodynamic erasure cost of a single bit with erasure error $1/2-\delta$. To calculate the work associated with the erasure of an ML bit, we first introduce the coordinate transformation
\begin{equation}
    \Gamma_\textrm{sb}(x)=f_{\textrm{sb},\tau}^{-1}(f_{\textrm{sb},0}(x)) \,,
\label{eq:gamdef}
\end{equation}
where $f_{\textrm{sb},0/\tau}(x)$ is the cumulative distribution associated with the state of the sub-bit at time $0$ or $\tau$, $p_{\textrm{sb},0/\tau}(x)$, respectively. Within the mathematical framework of optimal transport theory, $\Gamma(x)$ is known as the \textit{transport map}. Physically, it can be interpreted as the position to which one moves the probability density that was originally at position $x$. If $\delta=0$, the final macroscopic state of each sub-bit bit equals its initial state, $P_0=P_1=1/2$, {and the final microscopic state of the sub-bits is the same} as the distribution $p_{\textrm{sb},\tau}(x)=p_{\textrm{sb},0}(x)$. This implies that $\Gamma_\textrm{sb}(x)=x$. Therefore, in the limit of many sub-bits, one can Taylor expand $\Gamma_\textrm{sb}(x)$ to first order,
\begin{equation}
    \Gamma_{\textrm{sb}}(x) = x + \frac{\Gamma_{1}(x)}{\sqrt{N}} 
        + \order{\frac{1}{N}} \,,
\label{eq:gamdefmld}
\end{equation}
where $\Gamma_{1}(x)$ depends on the erasure protocol and can be interpreted physically as the total distance that the probability density originally at position $x$ is transported by time $\tau$.  In the appendix, we use Eq.~\eqref{eq:W} to show that one can now write
\begin{align}
    W_\textrm{min;ML} &= NW_{\textrm{min;sb}}\nonumber\\
    &\approx \int^{\infty}_{-\infty} \dd{x} \left( \frac{\left(p_0(x)\Gamma'_1(x)+p_0'(x)\Gamma_1(x)\right)^2}
    {2p_0(x)}\right. \nonumber \\
    &\hspace{1.5in}{\left.+\frac{p_0(x)\Gamma_1(x)^2}{D\tau}\right)} \,. 
\label{eq:wminmld}
\end{align}
Furthermore, the constraint $f_{\textrm{sb},\tau}(0)=1/2+\delta$ can be rewritten as
\begin{equation}
    \Gamma_1(0)=-\frac{\erf^{-1}(1-2\epsilon)}
        {\sqrt{2}p_0(0)} \,.
\label{eq:consmld}
\end{equation}

To calculate the minimum average work needed to erase an ML bit with erasure error $\epsilon$, one can minimize Eq.~\eqref{eq:wminmld} over all $\Gamma_{1}(x)$ that satisfy Eq.~\eqref{eq:consmld}. Lagrange's method of constrained optimization then leads to,
\begin{equation}
    \dv{x} \frac{\left(V'(x)\Gamma_1(x)\right)}{k_\textrm{B}T}-
    \dv[2]{x} \Gamma_1(x) + \frac{2}{D\tau} \Gamma_1(x)
    =\lambda\delta(x) \,,
\label{eq:gammamld}
\end{equation}
where $\lambda$ is a Lagrange multiplier.
As Eq.~\eqref{eq:gammamld} is a linear differential equation, once the initial potential-energy landscape of the sub-bits, $V(x,0)$, is specified, one can in principle determine analytically $\Gamma_1(x)$ and therefore the minimum work to erase an ML bit with many sub-bits. 

\begin{figure}
    \centering
    \includegraphics[width=3.25in]{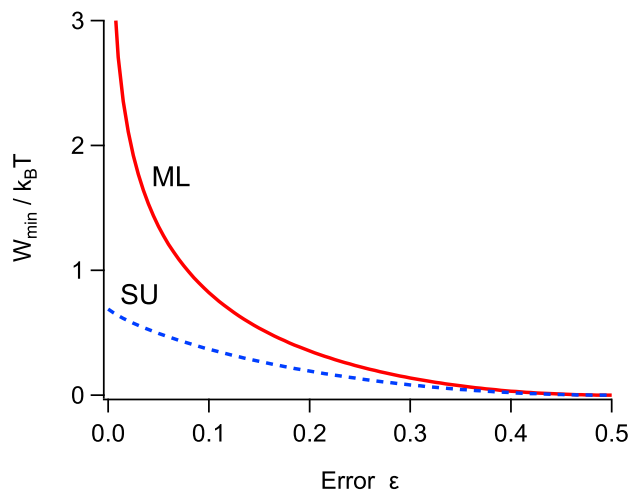}
    \caption{Minimum average work $W_{\textrm{min}}$ for a single-unit bit (blue dashed line) and a majority-logic bit (solid red line) in the slow-erasure limit.}
\label{fig:MLDslow}
\end{figure}
In the slow- and fast-erasure limits, one can calculate $W_\textrm{min;ML}$ exactly, even without specifying the $V(x,0)$. Indeed, in the slow-erasure, the work needed to erase the bit just reduces to the total change in Shannon entropy,
\begin{equation}
    W_\textrm{min;ML} \approx k_\textrm{B}T (\erf^{-1}(1-2\epsilon))^2 \,.
\label{eq:wminslow}
\end{equation}
In Fig.~\ref{fig:MLDslow}, we compare the slow-erasure costs of SU bit, Eq.~\eqref{eq:slowbound} with that of an ML bit, Eq.~\eqref{eq:wminslow}. One can immediately see that it always takes more work to erase an ML bit than an SU bit, whatever the erasure error.  In the fast-erasure limit, a WKB calculation leads to
\begin{equation}
    W_\textrm{min;ML} \approx \frac{k_\textrm{B}T (\erf^{-1}(1-2\epsilon))^2}
    {p_0(0)\sqrt{2D\tau}} \,.
\label{eq:wminfast}
\end{equation}
In the fast-erasure limit, the work needed to erase an ML bit scales as $\tau^{-1/2}$; by contrast, this cost for an SU bit scales as $\tau^{-1}$,c.f., Eq.~\eqref{eq:fast}. Therefore, in the fast-erasure limit, the cost associated with an SU bit scales quadratically worse than the cost associated with an ML bit.

\subsection{Finite-\textit{N} effects}

To arrive at the above results, Eq.~\ref{eq:gammamld}, we first took the limit $N \to \infty$ and then the limit $\tau \to 0$. Any realistic ML bit does, however, have a finite number of sub-bits. Therefore, for very fast erasure, the limit $\tau \to 0$ becomes the dominant limit and should be taken prior to the $N \to \infty$ limit. Using Eq.~\eqref{eq:fast}, one then arrives at
\begin{equation}
    W_\textrm{min;ML} \approx \frac{k_\textrm{B}T(\erf^{-1}(1-2\epsilon))^3}
        {\sqrt{72 N}p_0(0)^2D\tau} \,.
\label{eq:fintiteN}
\end{equation}
Comparing this result with Eq.~\eqref{eq:wminfast}, one can see that the finite-$N$ scaling becomes relevant for $\tau \sim \left( Dp_0(0)^2N \right)^{-1}$. Furthermore, by comparing Eq.~\eqref{eq:fintiteN} with Eq.~\eqref{eq:fast}, one can see that the ML bit can still outperform an SU bit by a factor $\sqrt{N}$.

\section{Example: Flat well \label{sec:ex}}
\begin{figure}
    \centering
    \includegraphics[width=3.25in]{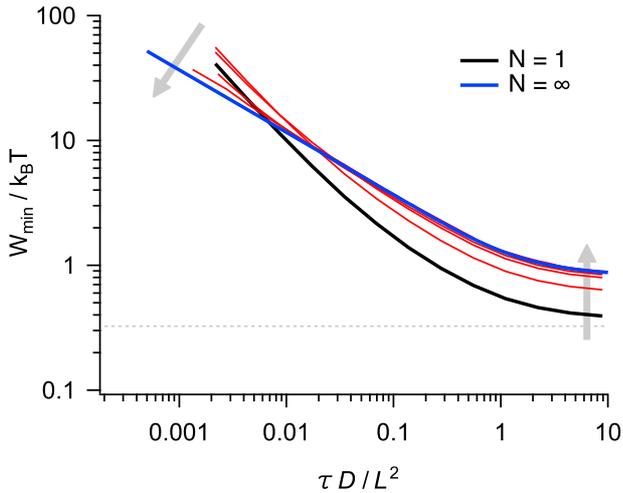}
    \caption{Minimum average work $W_{\textrm{min;ML}}$ to erase an ML bit as a function of scaled protocol duration $\tau' = \tau D/L^2$, for ML bits with $\epsilon=0.1$.  The thick black curve corresponds to $N=1$ sub-bit.  The thin red curves show $N=3,9,33,99$, and the thick black curve is the $N\to \infty$ result given in Eq.~\eqref{fig:MLDex}.  Light-shaded arrows show trends as $N$ is increased from 1 to $\infty$. 
    }
\label{fig:MLDex}
\end{figure}
Let us now apply our framework to a simple example, where we assume a large number of sub-bits ($N \to \infty$) and where the initial potential has a flat bottom of width $2L$ and infinite walls,
\begin{equation}
    V(x,0) =
        \begin{cases}
            0 & -L < x < L \\
            \infty & \textrm{otherwise}
        \end{cases} \,.
\label{eq:Vdef}
\end{equation}
To calculate the minimum work needed to erase the bit, we first need to find $\Gamma_{1}(x)$ from Eq.~\eqref{eq:gammamld}, with $V(x,0)$ given by Eq.~\eqref{eq:Vdef}. From Eqs.~\eqref{eq:gamdef} and \eqref{eq:gamdefmld}, it is clear that the solution should satisfy the boundary conditions
\begin{equation}
    \Gamma_1\left(-L\right) = 
    \Gamma_1\left(L\right)=0 \,.
\end{equation}
{Together with the constraint provided by Eq.~\eqref{eq:consmld}, this} leads to the solution (see appendix)
\begin{equation}
    \Gamma_1(x) =    
        \begin{cases}
            -\frac{\sqrt{2}L\erf\left(1-2\epsilon\right)\sinh\left(\sqrt{\frac{2}{\tau'}}\left(\frac{x}{L}+1\right)\right)}
            {\sinh\left(\sqrt{\frac{2}{D\tau}}L\right)} & x<0 \,,
                \\[15pt]
            -\frac{\sqrt{2}L\erf\left(1-2\epsilon\right)\sinh\left(\sqrt{\frac{2}{\tau'}}\left((\frac{x}{L}-1\right)\right)}
            {\sinh\left(-\sqrt{\frac{2}{\tau'}}\right)} & x>0 \,.
        \end{cases}
\label{gamex}
\end{equation}
Plugging this expression for $\Gamma_1(x)$ into Eq.~\eqref{eq:wminmld} then gives us the minimum work to erase an ML bit made up of a large number of sub-bits that each have an initial flat-well potential described by Eq.~\eqref{eq:Vdef}:
\begin{equation}
    W_{\textrm{min;ML}} = 
    \frac{k_\textrm{B}T \, (\erf^{-1}(1-2\epsilon))^2}{\sqrt{2\tau'}}
    \frac{\sinh\left(2\sqrt{\frac{2}{\tau'}}\right)}
    {\cosh^2\left(\sqrt{\frac{2}{\tau'}})\right)-1} \,,
\end{equation}
with $\tau' \equiv \tau D / L^2$ the scaled protocol time.

In Fig.~\ref{fig:MLDex}, we plot the minimum work needed to erase an ML bit, for various numbers of sub-bits. For large $\tau$, the cost to erase an SU bit is lower than the cost to erase an ML bit; however, for $\tau \lesssim\textrm{Var}(x)/D$, ML bits are easier to erase. In other words, an optimal protocol would use a single unit for slow erasure and majority logic for fast erasure. This conclusion that the optimal strategy shifts suddenly as a function of a parameter (the number of sub-bits $N$) recalls the scenarios of phase-transitions in optimal protocols studied in \cite{Solon2018}.  One can also see that for very small $\tau$, the finite-size effects of ML bits become important, making them more costly to erase.

\section{Discussion\label{sec:con}}

Why are single-unit bits more efficient for slow erasure but majority-logic bits more efficient for fast erasure?  In an ML bit, one neglects some microscopic degrees of freedom.  As a result, both the entropy and free-energy difference between the initial and final states of an ML bit exceed those of an SU bit. For fast erasure, the dissipative part of the work needed to erase the bit dominates. Following Eq.~\eqref{gamex} and interpreting $\Gamma_{\textrm{sb},1}(x)$ as the displacement of the probability distribution, one can see that optimal fast erasure {for} ML bits consists of changing the states only for sub-bits where $x\approx 0$. Meanwhile, sub-bits that have $x\gg 0$ and are therefore harder to erase remain untouched. The lower cost of sub-bits with $x\approx 0$ also explains why the erasure cost of an ML bit is inversely proportional to the probability that $x\approx 0$, as suggested by Eq.~\eqref{eq:wminfast}.

This conclusion points to an important design principle for realistic ML bits: Generally, one does not have full control over the energy landscape of the sub-bits, {and control is limited to the possibilities provided by a finite number of parameters.  Limiting the space of control algorithms will increase the} minimum work given by Eq.~\eqref{eq:wminfast}. The above argument does, however, show that the main ingredient to reduce the operational cost of an ML bit is the possibility to modify the energy landscape around $x=0$ without modifying it away from this region.  {That kind of manipulation of the potential can be created using a limited set of control parameters.}

It is worth noting that other effects might also lead to sublinear scaling of the work cost in the fast-erasure limit. For example, Pancotti et al.~have recently suggested that a time-dependent system-bath coupling might lead to similar effects \cite{pancotti2020speed}. Furthermore, effects inside the bath, such as viscoelasticity, might influence the erasure cost~\cite{zhang2019work}.  These mechanisms are quite distinct from the one proposed here, which has the advantage of applying to the types of memories used in technological applications, where a single bit consists of many spins and the average magnetization records the bit of information.

Can one extend our analysis to other types of systems? Optimal transport theory predicts that the minimum dissipation to transform a microscopic system from an initial to a final state is inversely proportional to the duration of the protocol. Here, we showed that if the mapping from microscopic to macroscopic dynamics is non-trivial, this conclusion need not hold for macroscopic transformations. {Similar analyses might then be possible in other settings, such as chemical reaction networks or heat engines.}


\acknowledgments
We thank Jannik Ehrich and Massimiliano Esposito for discussions.  This research was supported by a Natural Sciences and Engineering Research Council of Canada (NSERC) Discovery Grant and also by grant number FQXi-IAF19-02 from the Foundational Questions Institute Fund, a donor-advised fund of the Silicon Valley Community Foundation.  KP received funding as a postdoctoral fellow of the Research Foundation-Flanders (FWO).

\section{Appendix: erasure error\label{sec:app1}}
Here, we find the erasure error of an ML bit given the erasure errors of its sub-bits. First, we calculate the probability, $P_0$, that an MLB with $N$ sub-bit is in state $0$ when all of the sub-bits are in state $0$ with probability $p$.

It was shown in \cite{sheng2019thermodynamics} that
\begin{equation}
    P_0 = \frac{\int^p_0 \dd{t} \left( t(1-t) \right)^{\frac{N-1}{2}}}
    {\int^1_0 \dd{t} \left( t(1-t) \right)^{\frac{N-1}{2}}} \,.
\label{eq:P0}
\end{equation}
Let us now look at the large-$N$ limit, where the integrands in both the numerator and denominator of Eq.~\eqref{eq:P0} are exponentially small unless $t\approx 1/2$.  We can then do a saddle-point approximation:
\begin{align}
    P_0
    &\approx \frac{\int^p_0 \dd{t} \e^{-N\left(\ln 2 + 4 \left( t-\frac{1}{2} \right)^2 \right)}}
    {\int^1_0 \dd{t} \e^{-N\left(\ln 2 + 4 \left( t-\frac{1}{2} \right)^2 \right)}} \nonumber \\
    &\approx \frac{1}{2} \left[ 1+\erf\left(\sqrt{N}(2p-1)\right) \right] \,,
\label{eq:Pmld}
\end{align}
where we used the Taylor expansion of the log term around $t=1/2$,
\begin{equation}
    \ln\left(t(1-t)\right)\approx-2\ln 2-4\left(t-\frac{1}{2}\right)^2+\order{\left(t-\frac{1}{2}\right)^4}
\end{equation}
to get the second equality. Note that when $\sqrt{N}(p-1/2)\ll 0$ or $\sqrt{N}(p-1/2)\gg 0$, one gets $P_0\approx 0$ or $P_0\approx 1$ respectively, up to an exponentially small correction.

\section{Appendix: optimal work\label{sec:app2}}
Let us look at the large-$N$ limit and assume that $\Gamma_\textrm{sb}(x)$ is of the form
\begin{equation}
    \Gamma_\textrm{sb}(x) = x + \frac{\Gamma_1(x)}{\sqrt{N}} + \order{\frac{1}{N}} \,.
\label{eq:gammasb}
\end{equation}
The dissipation associated with the erasure of one sub-bit is given by

\begin{align}
    \Delta_\textrm{i}S_{\textrm{min;sb}} 
    &= \frac{\int^{1}_{0} \dd{y}
    \left(f^{-1}_{\textrm{sb},0}(y)-f^{-1}_{\textrm{sb},\tau}(y)\right)}{D\tau}\nonumber\\[3pt]
    &= \frac{\int^{\infty}_{-\infty} \dd{x} p_0(x) \left( \Gamma_\textrm{sb}(x)-x \right)}{D\tau} \nonumber \\[3pt]
    &\approx \int^{\infty}_{-\infty} \dd{x} \frac{p_0(x)\Gamma_1(x)^2}{ND\tau} \,,
\end{align}
where we changed variables from $x\to f_0(x)$ in the second line.  For the nonequilibrium free-energy difference, we first note that as $f_{\sbb,0}\left(x\right)=f_{\sbb,\tau}\left( \Gamma_{\sbb}(x)\right)$, we can write,
\begin{equation}
    p_{\sbb,0}(x) = \Gamma'_\sbb(x)p_{\sbb,\tau}(\Gamma_\sbb(x)) \,.
\end{equation}
This leads to
\begin{align}
   \frac{\Delta \mathcal{F}}{k_\textrm{B}T}
    &= \int^{\infty}_{-\infty} \dd{x} p_0(x) \ln \frac{p_0(x)}{\Gamma'(x)p_0(\Gamma(x))} \,,
\end{align}
where we changed variables, $x\to \Gamma(x)$ in the second line.  Plugging in Eq.~\eqref{eq:gammasb} and doing a lowest-order expansion in $1/\sqrt{N}$ leads to
\begin{equation}
    \frac{\Delta \mathcal{F}}{k_\textrm{B}T}=\frac{1}{N}\int^{\infty}_{-\infty} \dd{x} \left( \frac{\left( \dv{x} \left(p_0(x)\Gamma_1(x)\right) \right)^2}{2p_0(x)}\right) \,,
\end{equation}
and therefore we arrive at
\begin{equation}
    \frac{W_{\textrm{min;ML}} }{k_\textrm{B}T}= \int^{\infty}_{-\infty}  \dd{x} \left( \frac{\left( \dv{x} \left(p_0(x)\Gamma_1(x)\right)\right)^2}{2p_0(x)}+\frac{p_0(x)\Gamma_1(x)^2}{D\tau}\right).
\label{eq:wmin}
\end{equation}

Now let us reformulate the constraint
\begin{equation}
    f_{\sbb,\tau}(0)=\frac{1}{2}+\delta,\quad \textrm{or} \qquad \Gamma_{\sbb}\left(f_{\sbb,0}^{-1}\left(\frac{1}{2}+\delta\right)\right) = 0
\end{equation}
in terms of $\Gamma_1(x)$. First, we note that $f_0^{-1}(1/2)=0$. We then use a first-order Taylor expansion and Eq.~\eqref{eq:Pmld} to show that
\begin{align}
    f_{\sbb,0}^{-1}\left(\frac{1}{2}+\delta\right)
    &= \frac{\delta}{f_{\sbb,0}'\left(f_{\sbb,0}^{-1} \left( \frac{1}{2}\right)\right)}+O(\delta^2) \nonumber \\
    &= \frac{\erf^{-1}(1-2\epsilon)}{p_0(0)\sqrt{2N}}
    +\order{\frac{1}{N}} \,.
\end{align}
This leads to
\begin{equation}
    \Gamma_{\sbb}\left(f_{\sbb,0}^{-1}\left(\frac{1}{2}+\delta\right)\right) \approx \frac{\erf^{-1}(1-2\epsilon)}{p_0(0)\sqrt{2N}} + \frac{\Gamma_1(0)}{\sqrt{N}}=0 \,,
\end{equation}
or
\begin{equation}
    \Gamma_1(0) = -\frac{\erf^{-1}(1-2\epsilon)}
        {\sqrt{2}p_0(0)} \,,
\label{eq:bc}
\end{equation}
which can be rewritten to
\begin{equation}
    \int_{-\infty}^{\infty} \dd{x} \Gamma_1(x)\delta(x) = -\frac{\erf^{-1}(1-2\epsilon)}{\sqrt{2}p_0(0)} \,.
\label{eq:consint}
\end{equation}

The minimization of Eq.~\eqref{eq:wmin} under the constraint of Eq.~\eqref{eq:consint} can now be done by introducing the Lagrangian
\begin{eqnarray}
    \mathcal{L} =& \int^{\infty}_{-\infty} \dd{x} \left( \frac{\left( \dv{x} \left(p_0(x)\Gamma_1(x)\right)\right)^2}
    {2p_0(x)}+\frac{p_0(x)\Gamma_1(x)^2}{D\tau} \right) \nonumber\\&+\lambda\left(\int_{-\infty}^{\infty} \dd{x} \Gamma_1(x)\delta(x)+\frac{\erf^{-1}(1-2\epsilon)}{\sqrt{2}p_0(0)}\right)
\end{eqnarray}
and solving the Euler-Lagrange equation
\begin{equation}
    \dv{x} \fdv{\mathcal{L}}{\Gamma_1'(x)} 
    =\fdv{\mathcal{L}}{\Gamma_1(x)} \,,
\end{equation}
which leads to
\begin{multline}
    \dv{x} \left(p_0(x)\Gamma'_1(x)\right)+p_0''(x)\Gamma_1(x)-\frac{p_0'(x)^2\Gamma_1(x)}{p_0(x)}\\-\frac{2p_0(x)\Gamma_1(x)}{D\tau}=\lambda\delta(x),
\end{multline}
or using the fact that the system was initially in a Boltzmann distribution,
\begin{equation}
    \frac{\dv{x} \left(\Gamma_1(x)V'(x)\right)}{k_\textrm{B}T}-\dv[2]{x} \Gamma_1(x)+\frac{2\Gamma_1(x)}{D\tau}=\lambda\delta(x)
\label{eq:consmld2}
\end{equation}

\section{Appendix: slow- and fast-erasure limits\label{sec:app3}}

For slow erasure ($\tau \to \infty$), the minimum work needed to erase a bit is dominated by the free-energy difference between the initial and final states of the system,
\begin{align}
    \frac{W_{\textrm{min;ML}}}{k_\textrm{B}T} 
    &\approx \frac{\Delta F}{k_\textrm{B}T} \nonumber \\
    &= -N\left[] \epsilon_\textrm{sb} \ln \epsilon_\textrm{sb} + (1-\epsilon_\textrm{sb})\ln(1-\epsilon_\textrm{sb})
    -\ln 2 \right] \,,
\end{align}
where the second line follows from the fact that the local equilibrium distribution,
\begin{equation}
    p_\tau(x)=\begin{cases} 2(1-\epsilon)p_0(x) & x<0\\
    2\epsilon p_0(x) & x>0
    \end{cases}
\end{equation}
minimizes the free-energy difference between the initial and final states \cite{proesmans2020optimal}. Using, $\epsilon_\textrm{sb}=1/2-\delta$ with $\delta$ small, we can rewrite this as
\begin{align}
     W_{\textrm{min;ML}} &\approx 2Nk_\textrm{B}T \delta^2 
        \nonumber \\ 
    &= k_\textrm{B}T\erf^{-1}\left(1-2\epsilon\right)^2.
\end{align}

{For fast erasure ($\tau \to 0$),} we first need to solve Eq.~\eqref{eq:consmld2}. To do this, we first make a WKB ansatz for $\Gamma_1(x)$,
\begin{equation}
    \Gamma_1(x)\sim\exp\left(\frac{\alpha_0(x)}{\sqrt{\tau}}+\alpha_1(x)+O(\sqrt{\tau})\right).
\end{equation}
Eq.~\eqref{eq:consmld2} then becomes
\begin{multline}
    \frac{2-D\alpha_0'(x)^2}{D\tau}+\frac{a_0'(x)\left(\frac{V'(x)}{k_\textrm{B}T}-2a_1'(x)\right)-a_0''(x)}{\sqrt{\tau}}\\+O(1)=\lambda\delta(x).
\end{multline}
We can now approximate $\Gamma_1(x)$ by solving the above equation order-by-order in $\tau$:
\begin{align}
    2-D\alpha_0'(x)^2 &= 0 \\
    a_0'(x) \left( V'(x)-2a_1'(x) \right) - a_0''(x) &= 0 \,.
\end{align}
This leads to
\begin{equation}
    \alpha_0(x) = \pm\sqrt{\frac{2}{D}}x,\qquad 
    \alpha_1(x) = \frac{V(x)}{2k_\textrm{B}T} \,.
\end{equation}
Finally, as Eq.~\eqref{eq:consmld} is a homogeneous linear differential equation for $x\neq 0$, any linear combination of solutions is itself a solution. The delta function has the effect that the first derivative of $\Gamma_1(x)$ need not be continuous at $x\neq 0$. Therefore, the general solution can be written as
\begin{equation}
    \Gamma_1(x) = 
    \begin{cases}
        \e^{\frac{V(x)}{2k_\textrm{B}T}}\left(C_{0,-} \e^{\sqrt{\frac{2}{D\tau}}x}+C_{1,-} \e^{-\sqrt{\frac{2}{D\tau}}x}\right) & x<0 \,, \\
        \e^{\frac{V(x)}{2k_\textrm{B}T}}\left(C_{0,+} \e^{\sqrt{\frac{2}{D\tau}}x} + C_{1,+} \e^{-\sqrt{\frac{2}{D\tau}}x}\right) & x>0 \,.
    \end{cases}
\end{equation}
One can now fix $C_{0,-}$, $C_{1,-}$, $C_{0,+}$ and $C_{1,+}$ by using the boundary conditions, Eq.~\eqref{eq:bc}, and $\Gamma_1(x_\textrm{min})=\Gamma_1(x_\textrm{max})=0$.   Using these conditions, we can write
\begin{equation}
    \Gamma_1(x) =
    \begin{cases}
        \frac{\erf^{-1}(1-2\epsilon)}{\sqrt{2}p_0(0)} \e^{\frac{V(x)-V(0)}{2k_\textrm{B}T}}\frac{\sinh(\sqrt{\frac{2}{D\tau}}
        (x-x_\textrm{min}))}{\sinh(\sqrt{\frac{2}{D\tau}}x_\textrm{min})} & x<0 \,, \\
    \frac{\erf^{-1}(1-2\epsilon)}{\sqrt{2}p_0(0)} \e^{\frac{V(x)-V(0)}{2k_\textrm{B}T}}\frac{\sinh(\sqrt{\frac{2}{D\tau}}{(x-x_\textrm{max})})}{\sinh(\sqrt{\frac{2}{D\tau}}{x_\textrm{max}})} & x>0 \,.
    \end{cases}
\end{equation}
For small $\tau$, it is clear that $\Gamma_1(x)$ is exponentially small unless $\left| x\right|\approx 0$, where one can write
\begin{equation}
    \Gamma_1(x)\approx \frac{\erf^{-1}(1-2\epsilon)}{\sqrt{2}p_0(0)} \e^{-\sqrt{\frac{2}{D\tau}}\left| x\right|} \,.
\end{equation}
Therefore, we can do a saddle-point-like approximation on Eq.~\eqref{eq:wmin}, which leads to
\begin{align}
    \frac{W_{\textrm{min;ML}}}{k_\textrm{B}T}\approx& \left( \int^{\infty}_{-\infty} \dd{x}\frac{\left(p_0'(0) \e^{-\sqrt{\frac{2}{D\tau}}\left|x\right|}-p_0(0)\sqrt{\frac{2}{D\tau}} \e^{-\sqrt{\frac{2}{D\tau}}\left|x\right|}\right)^2}
    {2p_0(0)}\right.\nonumber\\&\left.+\int^{\infty}_{-\infty} \dd{x}\frac{p_0(0) \e^{-2\sqrt{\frac{2}{D\tau}}\left|x\right|}}{D\tau} \right) \left(\frac{\erf^{-1}(1-2\epsilon)}{\sqrt{2}p_0(0)}\right)^2\nonumber\\[3pt]
    \approx& \frac{2\erf^{-1}(1-2\epsilon)^2}{p_0(0)D\tau}\int^{\infty}_{-\infty} \dd{x} \e^{-2\sqrt{\frac{2}{D\tau}}\left|x\right|} \nonumber \\[3pt]
    =& \frac{\sqrt{2}\erf^{-1}(1-2\epsilon)^2}{p_0(0)\sqrt{D\tau}} \,.
\label{eq:WminWKB}
\end{align}

\section{Appendix: flat well\label{sec:app4}}

As an example, let us look at a bit consisting of a particle in a flat potential well of width $2L$,
\begin{equation}
    V(x)=\begin{cases} 
        0 & -L<x<L \\
        \infty & \textrm{otherwise}
    \end{cases}
\end{equation}
The equilibrium distribution is given by
\begin{equation}
    p_0(x)=\frac{1}{2}L \,,
\end{equation}
for $-L<x<L$ and zero otherwise. Eq.~\eqref{eq:consmld} then becomes
\begin{equation}
    -\dv[2]{x} \Gamma_1(x) + \frac{2\Gamma_1(x)}{D\tau} = \lambda\delta(x) \,,
\end{equation}
with boundary conditions $\Gamma_1(-L)=\Gamma_1(L)=0$ and $\Gamma_1(0)=-\sqrt{2}l\erf^{-1}(1-2\epsilon)$. Therefore, one gets
\begin{equation}
    \Gamma_1(x) = 
    \begin{cases}         
        \sqrt{2}L \, {\erf^{-1}(1-2\epsilon)}
        \frac{\sinh(\sqrt{\frac{2}{D\tau}}(x+L))}{\sinh(\sqrt{\frac{2}{D\tau}}L)} & x<0 \\[9pt]
        -\sqrt{2}L \, {\erf^{-1}(1-2\epsilon)}
        \frac{\sinh(\sqrt{\frac{2}{D\tau}}{(x-L)})}
        {\sinh(\sqrt{\frac{2}{D\tau}}{L})} & x>0
    \end{cases}
\end{equation}
This leads to
\begin{align}
    W_{\textrm{min;ML}} &= \int^L_{-L} \dd{x} \left( \frac{\Gamma'_1(x)^2}{4L} +
    \frac{\Gamma_1(x)^2}{2LD\tau}\right) \nonumber\\
    &= \frac{k_\textrm{B}T L \, (\erf^{-1}(1-2\epsilon))^2}{\sqrt{2D\tau}} \,
    \frac{\sinh\left(2L\sqrt{\frac{2}{D\tau}}\right)}
    {\left(\cosh\left( \sqrt{\frac{2}{D\tau}}L)\right)^2-1 \right)} \,.
\label{eq:wminex}
\end{align}
\begin{figure}
    \centering
    \includegraphics[width=3.25in]{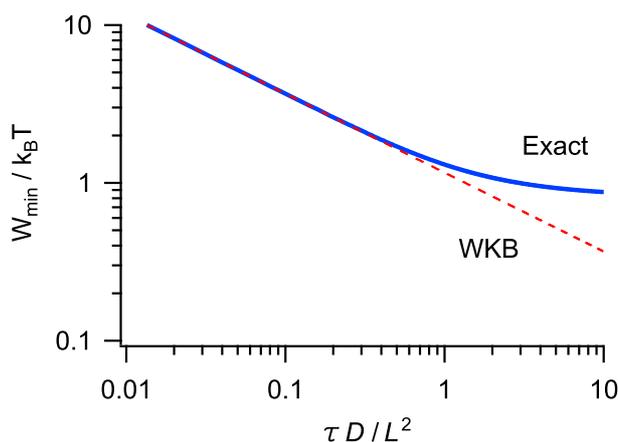}
    \caption{Minimum average work $W_{\textrm{min;ML}}$ to erase an ML bit as a function of scaled protocol duration $\tau' = \tau D/L^2$, for an ML bit with $\epsilon=0.1$.  The thick blue curve corresponds to the exact $N=\infty$ solution (Eq.~\ref{eq:wminex}).  The dashed red curve shows the WKB approximation from Eq.~\eqref{eq:WminWKB}, with $p_0(0) = \tfrac{1}{2}L$.}
\label{fig:MLDex2}
\end{figure}


\begin{thebibliography}{10}
\expandafter\ifx\csname url\endcsname\relax\def\url#1{\texttt{#1}}\fi

\bibitem{landauer1961irreversibility}
\Name{Landauer R.} \REVIEW{IBM J. Res. Dev.}{5}{1961}{183}.

\bibitem{koomey2010implications}
\Name{Koomey J., Berard S., Sanchez M. \and Wong H.} \REVIEW{IEEE Ann. Hist.
  Comput.}{33}{2010}{46}.

\bibitem{seifert2012stochastic}
\Name{Seifert U.} \REVIEW{Rep. Prog. Phys.}{75}{2012}{126001}.

\bibitem{van2015ensemble}
\Name{Van~den Broeck C. \and Esposito M.} \REVIEW{Physica A}{418}{2015}{6}.

\bibitem{dillenschneider2009memory}
\Name{Dillenschneider R. \and Lutz E.} \REVIEW{Phys. Rev.
  Lett.}{102}{2009}{210601}.

\bibitem{aurell2012refined}
\Name{Aurell E., Gawedzki K., Mejia-Monasterio C., Mohayaee R. \and
  Muratore-Ginanneschi P.} \REVIEW{J. Stat. Phys.}{147}{2012}{487}.

\bibitem{diana2013finite}
\Name{Diana G., Bagci G.~B. \and Esposito M.} \REVIEW{Phys. Rev.
  E}{87}{2013}{012111}.

\bibitem{zulkowski2014optimal}
\Name{Zulkowski P.~R. \and DeWeese M.~R.} \REVIEW{Phys. Rev.
  E}{89}{2014}{052140}.

\bibitem{zulkowski2015optimal}
\Name{Zulkowski P.~R. \and DeWeese M.~R.} \REVIEW{Phys. Rev.
  E}{92}{2015}{032117}.

\bibitem{deshpande2017designing}
\Name{Deshpande A., Gopalkrishnan M., Ouldridge T.~E. \and Jones N.~S.}
  \REVIEW{Proc. R. Soc. Lond.}{473}{2017}{20170117}.

\bibitem{melbourne2018realizing}
\Name{Melbourne J., Talukdar S. \and Salapaka M.~V.} \REVIEW{Proc. IEEE Conf.
  on Decision and Control (CDC)}{}{2018}{4135}.

\bibitem{boyd2018shortcuts}
\Name{Boyd A.~B., Patra A., Jarzynski C. \and Crutchfield J.~P.}
  \REVIEW{arXiv:1812.11241}{}{2018}{}.

\bibitem{riechers2019balancing}
\Name{Riechers P.~M., Boyd A.~B., Wimsatt G.~W. \and Crutchfield J.~P.}
  \REVIEW{Phys. Rev. Research}{2}{2020}{033524}.

\bibitem{proesmans2020finite}
\Name{Proesmans K., Ehrich J. \and Bechhoefer J.} \REVIEW{Phys. Rev.
  Lett.}{125}{2020}{100602}.

\bibitem{proesmans2020optimal}
\Name{Proesmans K., Ehrich J. \and Bechhoefer J.} \REVIEW{Phys. Rev.
  E}{102}{2020}{032105}.

\bibitem{berut2012experimental}
\Name{B{\'e}rut A., Arakelyan A., Petrosyan A., Ciliberto S., Dillenschneider
  R. \and Lutz E.} \REVIEW{Nature}{483}{2012}{187}.

\bibitem{berut2013detailed}
\Name{B{\'e}rut A., Petrosyan A. \and Ciliberto S.}
  \REVIEW{EPL}{103}{2013}{60002}.

\bibitem{jun2014high}
\Name{Jun Y., Gavrilov M. \and Bechhoefer J.} \REVIEW{Phys. Rev.
  Lett.}{113}{2014}{190601}.

\bibitem{gavrilov2016erasure}
\Name{Gavrilov M. \and Bechhoefer J.} \REVIEW{Phys. Rev.
  Lett.}{117}{2016}{200601}.

\bibitem{hong2016experimental}
\Name{Hong J., Lambson B., Dhuey S. \and Bokor J.} \REVIEW{Sci.
  Adv.}{2}{2016}{e1501492}.

\bibitem{martini2016experimental}
\Name{Martini L., Pancaldi M., Madami M., Vavassori P., Gubbiotti G., Tacchi
  S., Hartmann F., Emmerling M., H{\"o}fling S., Worschech L. \etal}
  \REVIEW{Nano Energy}{19}{2016}{108}.

\bibitem{gavrilov2017direct}
\Name{Gavrilov M., Ch{\'e}trite R. \and Bechhoefer J.} \REVIEW{Proc. Natl.
  Acad. Sci. U.S.A.}{114}{2017}{11097}.

\bibitem{gaudenzi2018quantum}
\Name{Gaudenzi R., Burzur{\'\i} E., Maegawa S., van~der Zant H. \and Luis F.}
  \REVIEW{Nat. Phys.}{14}{2018}{565}.

\bibitem{ribezzi2019large}
\Name{Ribezzi-Crivellari M. \and Ritort F.} \REVIEW{Nat. Phys.}{15}{2019}{660}.

\bibitem{saira2020nonequilibrium}
\Name{Saira O.-P., Matheny M.~H., Katti R., Fon W., Wimsatt G., Crutchfield
  J.~P., Han S. \and Roukes M.~L.} \REVIEW{Phys. Rev.
  Research}{2}{2020}{013249}.

\bibitem{cetiner2020dissipation}
\Name{Cetiner U., Raz O. \and Sukharev S.}
  \REVIEW{bioRxiv}{}{2020}{2020.06.26.174649;}.

\bibitem{aurell2011optimal}
\Name{Aurell E., Mej{\'\i}a-Monasterio C. \and Muratore-Ginanneschi P.}
  \REVIEW{Phys. Rev. Lett.}{106}{2011}{250601}.

\bibitem{schmiedl2008efficiency}
\Name{Schmiedl T. \and Seifert U.} \REVIEW{EPL}{83}{2008}{30005}.

\bibitem{richter1999recent}
\Name{Richter H.~J.} \REVIEW{J. Phys. D}{32}{1999}{R147}.

\bibitem{sheng2019thermodynamics}
\Name{Sheng S., Herpich T., Diana G. \and Esposito M.}
  \REVIEW{Entropy}{21}{2019}{284}.

\bibitem{Esposito2011NonequilibriumFreeEnergy}
\Name{Esposito M. \and Van~den Broeck C.} \REVIEW{EPL}{95}{2011}{40004}.

\bibitem{zhang2019work}
\Name{Zhang Y.} \REVIEW{EPL}{128}{2019}{30002}.

\bibitem{zhang2020optimization}
\Name{Zhang Y.} \REVIEW{J. Stat. Phys.}{178}{2020}{1336}.

\bibitem{Solon2018}
\Name{Solon A.~P. \and Horowitz J.~M.} \REVIEW{Phys. Rev.
  Lett.}{120}{2018}{180605}.

\bibitem{pancotti2020speed}
\Name{Pancotti N., Scandi M., Mitchison M.~T. \and Perarnau-Llobet M.}
  \REVIEW{Phys. Rev. X}{10}{2020}{031015}.

\end{thebibliography}
\end{document}